\documentclass[%
 reprint,
 amsmath,amssymb,
 aps,
prl,
]{revtex4-2}

\usepackage{graphicx}
\usepackage{dcolumn}
\usepackage{bm}
\usepackage{hyperref}
\usepackage{xcolor}
\begin{document}

\preprint{APS/123-QED}

\title{Chaotic scattering and heating in cold ion-atom collisions: two sides of the same coin}

\author{Saajid Chowdhury}
 \email{saajid.chowdhury@stonybrook.edu}
\author{Jesus Perez-Rios}%
 \email{jesus.perezrios@stonybrook.edu}
\affiliation{%
 Stony Brook University
}%

\date{\today}

\begin{abstract}
    We study the classical dynamics of a Paul-trapped ion in a low-density bath of atoms above $1\mu$K. We find that lower energy collisions with more massive atoms, especially at energies less than the initial micromotion heating, are more likely to form atom-ion complexes. These complexes evolve in a fractal structure for every scattering observable, showing non-hyperbolic chaotic dynamics. To explore the chaotic dynamics, we use a GPU-accelerated methodology allowing us to run over $3\times 10^8$ trajectories of a $^{174}\textrm{Yb}^+$ and different atoms. As a result, after analyzing the dynamics as a function of the atom species, collision energy, trap parameters, and ion-atom potential depth, we find a link between heating and the onset of chaos in the first atom-ion interaction that occurs when a low-density atomic bath is merged with a trapped ion. 
\end{abstract}

\maketitle


The pioneering work of Cetina et al.~\cite{Cetina2012} on the role of the trapping potential on ion-atom collisions served as a roadmap for cold chemistry and the hybrid ion-atom communities in the last decade~\cite{Tomza2019,LOUS202265,COTE20166}. The main idea is that as the ion moves away from the center of the trap as a consequence of its interaction with an atom, it experiences micromotion due to the time-dependent nature of the trapping potential. The ion displacement depends on the ion-to-atom mass ratio, as expected from kinematic considerations. Hence, a smaller ion displacement results in less micromotion heating. Therefore, to reach the lowest temperature possible for an ion in a neutral bath, it is optimal to choose a heavy ion and a light atom, such as the systems Yb$^+$-Li and Ba$^+$-Li. The former was the first to be close to the s-wave regime~\cite{Feldker2020}, and the latter has shown traces of ion-atom Feshbach resonances, indicating the onset of quantal control over ion-atom reactions~\cite{Weckesser2021,Thielemann2025}. However, even though it is well-known that the ion trap can induce heating, studies where the trap is considered are scarce~\cite{Hirzler2020,Hirzler2023,shi2025effectsdelocalizedchargedistribution,liang2025trapinducedatomioncomplexestimeindependent}

Only recently, it has been shown that time-dependent ion traps induce long-lived ion-atom complexes that can compromise the stability of the ion via three-body recombination reactions~\cite{Hirzler2023}, or can alter bimolecular processes~\cite{Pinkas2023}. Therefore, the trap can modify the outcome of ion-neutral reactions, depending on the nature of the reaction and temperature, even when the ion delocalization is included~\cite{shi2025effectsdelocalizedchargedistribution,liang2025trapinducedatomioncomplexestimeindependent}. Similarly, it has been shown that ion-atom complexes are a consequence of the onset of chaotic scattering in the system~\cite{Pinkas2024}. It has been predicted that an ion will show a larger probability of forming a complex in the case of a heavy atom bath than in a light one. Interestingly enough, in studies of a single trapped ion in a neutral bath, the very same conclusion is drawn regarding the ion heating rate: the heavier the atom is, the larger the ion heating~\cite{Furst2018,Trimby2022,Londono2022,Londono2023,Pinkas2020,BASTIAN2016,NIRANJAN2021}. Therefore, a priori, these independent conclusions indicate that the collision-induced heating of the ion (induced by the trap) and the formation of ion-atom complexes are correlated. The question is, how? The answer comes from treating the problem as a dynamical system, looking into the sensitive dependence of the scattering on initial conditions and parameters, to characterize the chaos, which ensues once the complex forms. 

In this Letter, we present a comprehensive study on ion-atom dynamics in the presence of a Paul trap, revealing that the onset of chaotic scattering is indeed triggered by the ion heating induced by a neutral bath. Our study is based on a GPU-accelerated framework, which enables us to simulate more than $3\times 10^8$ trajectories across different systems and trap parameters. As a result, we find that the probability of complex formation and ion heating rate are fully independent of the details of the short-range ion-atom interaction potential. Moreover, we find that the chaos in the system is non-hyperbolic, rather than hyperbolic as characteristic in scattering problems~\cite{Croft2014}, which has to do with the role of the time-dependent trap.

\begin{figure*}
    \includegraphics[width=\textwidth]{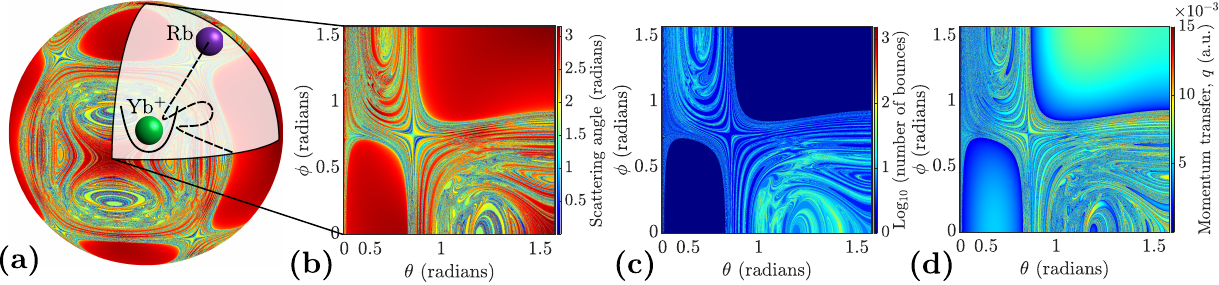}
    \caption{Chaotic structure of observables for head-on collisions of Yb$^+$Rb, for $D_e=1$K and $E=1.5\mu$K and $r_0=\textrm{5,000}a_0$, as a function of incoming atom's spherical angles $\theta$ and $\phi$. Panel (a) shows a schematic of the setup: the ion is placed at the center of the trap, and the atom in the surface of the octant of the sphere. The dashed line represents an ion-atom trajectory showing the formation of a complex. The heatmap of the scattering angle is projected on the surface of the sphere. Panel (b) shows the scattering angle, defined as $\arccos\frac{\vec{v}_i\cdot\vec{v}_f}{v_iv_f}$, where $\vec{v}_i$ and $\vec{v}_f$ are the initial and final atom velocities. Panel (c) shows the logarithm of the number of bounces in the atom-ion trajectory. Panel (d) shows the momentum transfer of the trajectory, defined as $q=|\vec{p}_\text{f}-\vec{p}_\text{i}|$, where $\vec{p}_\text{i}$ and $\vec{p}_\text{f}$ are the initial and final momentum of the atom. Due to outliers, $\textrm{min}(q,0.015\textrm{a.u.})$ is shown.} 
    \label{fig1}
\end{figure*}


Here, we simulate the trapped ion-atom scattering within a classical framework, which is fairly accurate down to temperatures of 1$\mu$K. The Hamiltonian of the system is 
\begin{align} 
    H=&\frac{1}{2}m_av_a^2+\frac{1}{2}m_iv_i^2+V_{ai}(|\vec{r}_a-\vec{r}_i|)+\\
      &\sum_{i=1}^3(a_i+2q_i\cos\Omega_\textrm{rf}t)\frac{m_\textrm{ion}\Omega_\textrm{rf}^2}{8}r_i^2
\end{align}
where $a_i$ and $q_i$ are parameters that depend on the trap and ion mass, and $\Omega_\textrm{rf}$ is the trap rf frequency. The interaction between the atom and ion is described by the potential $V_{ai}(r)=C_8/r^8-C_4/r^4$, where $C_4=\alpha/2$ and $\alpha$ is the polarizability of the atom in atomic units, and $C_8$ represents the short-range repulsion coefficient. To solve Newton's equations, we use a fourth/fifth order adaptive timestep Runge-Kutta method to propagate the differential equations for twelve variables: the positions and velocities of the ion of mass $m_i$ and atom of mass $m_a$, as a function of time. To increase the number of trajectories we could calculate, we implemented the Runge-Kutta algorithm in \cite{Dormand1980,Shampine1997} in a form that can be massively parallelized on a GPU \cite{Github}. This involved expressing the algorithm in terms of primitive operations and using scalar variables instead of vector variables, enabling automatic translation from MATLAB \cite{MATLAB} to CUDA. This allows us to run dozens of trajectories simultaneously per GPU, each trajectory running over 20x faster than MATLAB's \verb+ode45+ on a CPU, leading to a runtime more than 500x faster than running on a CPU, so that on one node with eight Tesla K80 24GB GPUs, we can simulate $10^7$ trajectories averaging 300,000 timesteps each in under 16 hours. 

We ran two different kinds of simulations, which we call head-on runs and thermal runs. For head-on runs, the ion starts at the center of the trap with zero velocity, and the atom starts on the surface of a large sphere of radius $r_0$, with a fixed velocity $v_0$ directed radially towards the ion at the center, fixed by the collision energy $E=\frac{1}{2}m_av_0^2$. For every simulation, the size of the sphere $r_0$ is chosen such that the condition $10|V(r_0)|<E$ is fulfilled, so that every collision starts in the asymptotic region where the atom motion is uniform rectilinear. 

Fig.~\ref{fig1} shows the results, for 10$^7$ trajectories for head-on Yb$^+$-Rb collisions at a collision energy of 1.5~$\mu$K, for the scattering angle, the number of bounces (short-range collisions) between the atom and ion, and the momentum transfer to the ion. Each of these scattering observables shows an identical pattern: a region of regular motion, where the atom simply scatters backwards from the ion after a single bounce, not forming a complex; and a fractal region, dominated by complex formation. This confirms the onset of chaotic scattering for trapped ion-atom collisions. The regular regions are initial points on the sphere from which incoming atoms are immediately scattered back out by the ion micromotion during the first encounter. The sizes and locations of these regions depend intricately on the trap parameters and the phase $\Omega_\textrm{rf}t$ mod $2\pi$ at moment of impact. The ``saddle point'' between the two regular regions occurs along $\phi=\pi/4$, dominated by complexes, since $q_x=-q_y$, while $x=y$, so the micromotion is perpendicular to the incoming atom velocity, always deflecting the ion laterally away from the approaching atom, stealing enough energy on the final largest micromotion cycle, to form a complex. It is worth emphasizing the fact that the momentum transfer to the ion [panel (d)], which quantifies the net heating of the ion from the beginning to the end of the complex trajectory, correlates with ion-atom complex formation, aligning with our hypothesis that ion heating and ion-atom complex formation are intimately related. 

\begin{figure}
    \includegraphics[width=0.5\textwidth]{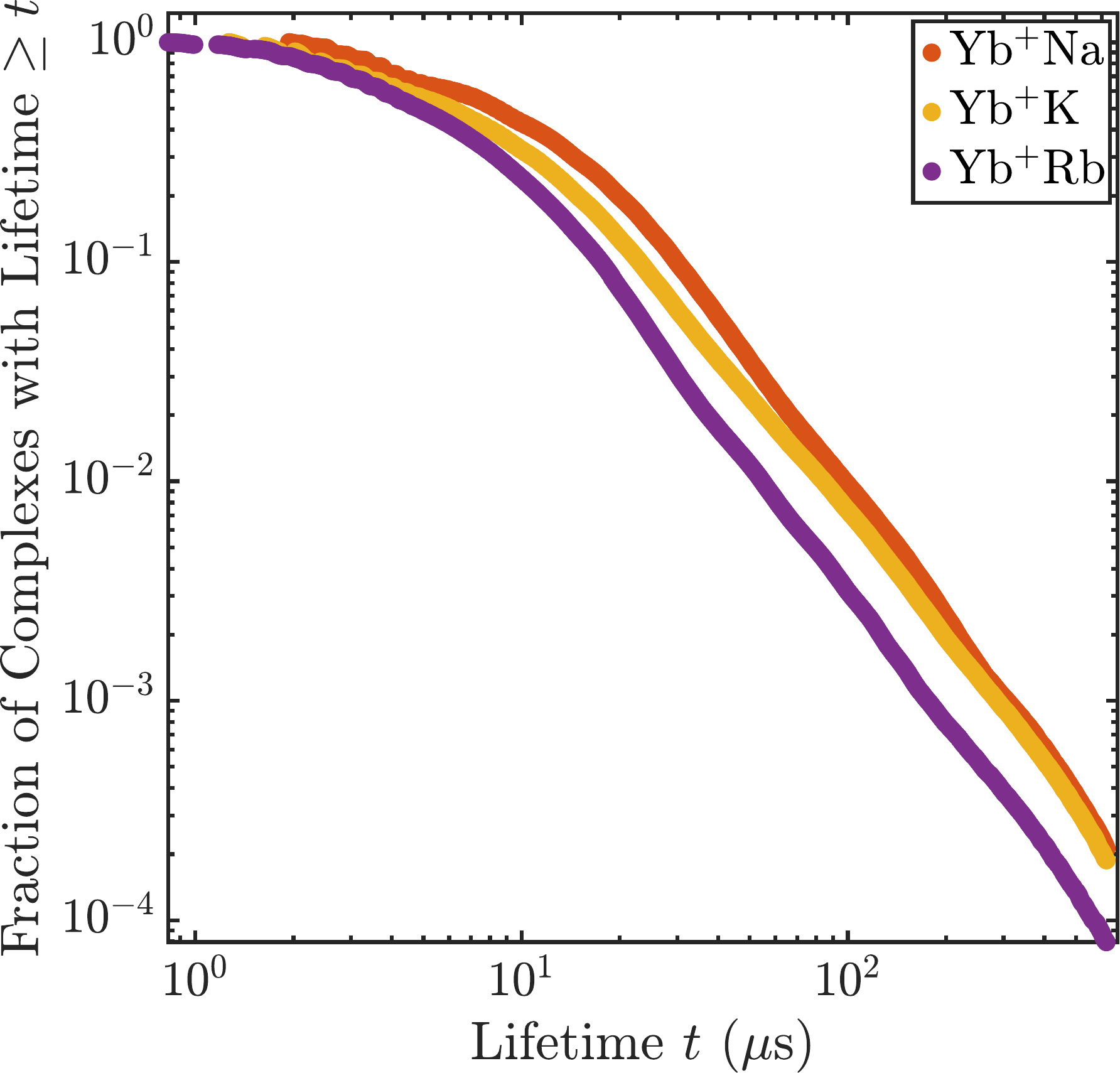}
    \caption{Fraction of complexes still alive after time $t$, for head-on collisions of different atoms with Yb$^+$ for $D_e=1$K and $E=1.5\mu$K and $r_0=\textrm{5,000}a_0$.} 
    \label{fig2}
\end{figure}

\begin{figure}
    \includegraphics[width=0.5\textwidth]{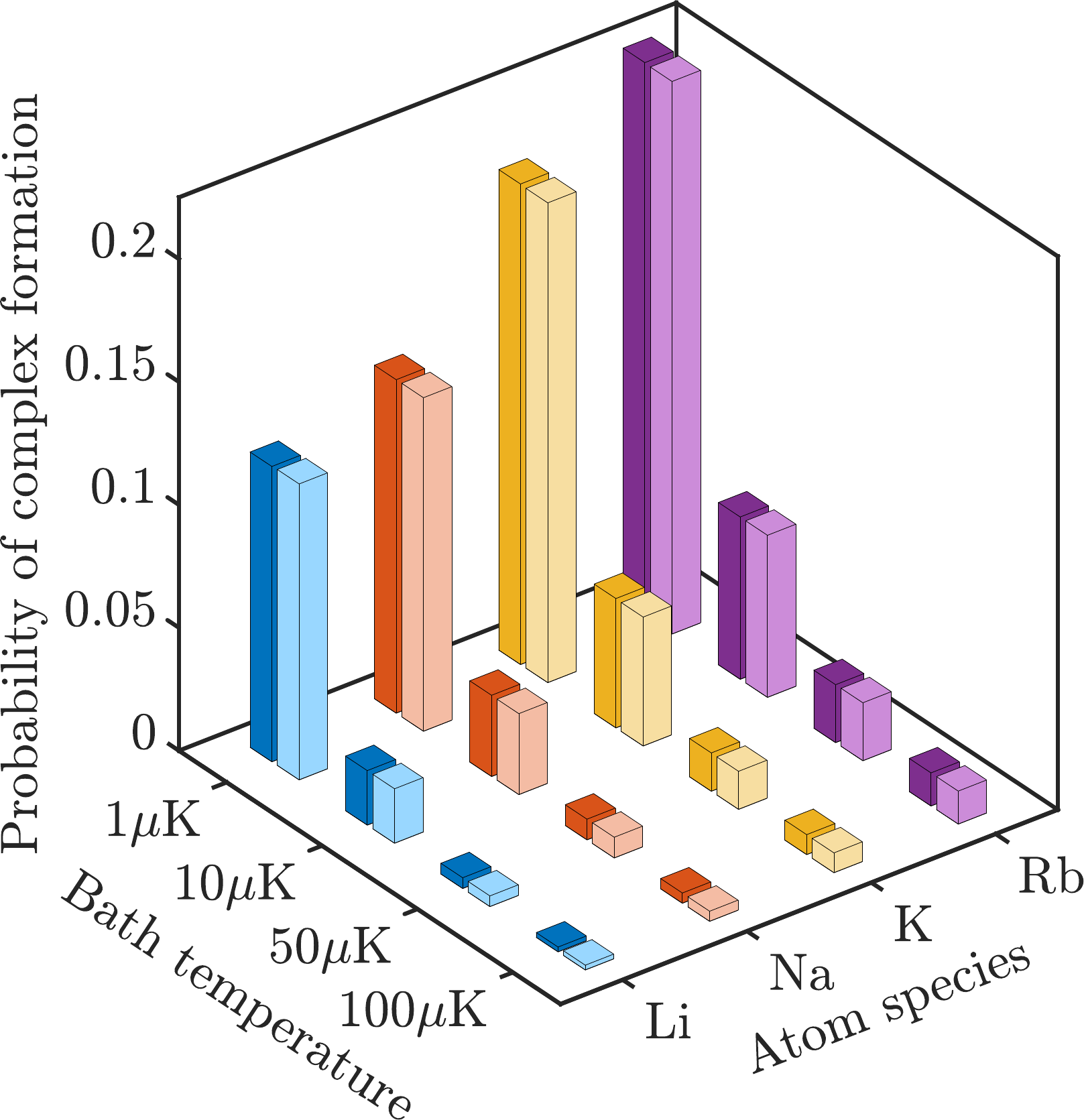}
    \caption{Probability of complex formation, as a function of bath temperature and atomic species, from thermal runs. The darker and lighter colors correspond to potential depths $D_e=100$K, $1$K respectively. Each bar represents the number of trajectories in which the atom and ion formed a complex, out of the $10^7$ computed trajectories. The ion starts near the center of the trap, and the atom starts from a uniformly chosen position on the surface of a sphere of radius 5,000$a_0$ centered at the trap center, with Cartesian velocity components $v_x,v_y,v_z$ sampled from a Gaussian distribution for the given bath temperature, negated if needed to enforce $v_r<0$.} 
    \label{fig3}
\end{figure}

To explore further the onset of chaotic scattering in trapped ion-atom collisions, we calculate the lifetimes of ion-atom complexes for different atomic species with the same ion. The results are shown in Fig.~\ref{fig2}, which shows the fraction of complexes remaining after time $t$, as a function of $t$. After about 20 $\mu$s, the lifetime distribution follows a power law. This is related to the fact that the time of first return of a one-dimensional random walk (e.g., the number of coin flips until equal numbers of heads and tails are obtained) has a distribution asymptotic to a power law \cite{Feller1968}; during each bounce, the trap can either add or subtract energy, until the trap adds enough energy to break the complex. This explanation is, however, an oversimplification, since it does not account for things such as the size of the bounce and the angular momentum exchange during the bounce. The power is not universal, as it depends on the atom, consistent with the probability dependence shown in Fig.~\ref{fig3}. This power law behavior shows that time-dependent-trapped ion-atom collision dynamics exhibit non-hyperbolic chaos, as opposed to hyperbolic chaos in scattering systems, which would show an exponential decay \cite{Croft2014,liang2025trapinducedatomioncomplexestimeindependent}. Verifying this trend up to lifetimes of over 600 $\mu$s relies on the great sampling of initial conditions that we count on in this work. 

For thermal runs, the ion's initial position and momentum are chosen from a Gaussian distribution corresponding to a fixed temperature of $10\mu$K for a spherical harmonic trap with frequency $\omega=q_x\Omega_\textrm{rf}/2^{3/2}$. For a given temperature $T$, the atom's initial velocity components $v_x,v_y,v_z$ were chosen from a Gaussian distribution from the Boltzmann weight $e^{-mv_i^2/2k_BT}$, and the atom's initial position was chosen uniformly on the surface of a sphere of radius $r_0=\textrm{5,000}a_0$ centered at the trap center. 

Since the chaotic region in Fig.~\ref{fig1} is dominated by complexes while the non-chaotic region has no complexes, we may use the probability of complex formation as a metric for how chaotic the system is. Thus, to investigate the physical conditions which give rise to chaotic dynamics, we conduct a comprehensive study of how the probability of complex formation depends on the temperature of the bath, the bath's atom species, and the short-range atom-ion physics. The results for over three hundred million trajectories in Fig.~\ref{fig3} show that heavier, cooler atoms are more likely to form complexes with an ion in a Paul trap, consistent with \cite{Hirzler2023}. Furthermore, when the atom-ion potential depth is changed by two orders of magnitude, there is no change in the probability, which shows that the short-range physics of the atom-ion interaction does not play a role in the chaotic aspects of the scattering. Thus, slower atoms spend more time attracting the ion away from the center of the trap, via the long-range interaction, which induces more micromotion heating and chaotic scattering. 

\begin{figure}
    \includegraphics[width=0.5\textwidth]{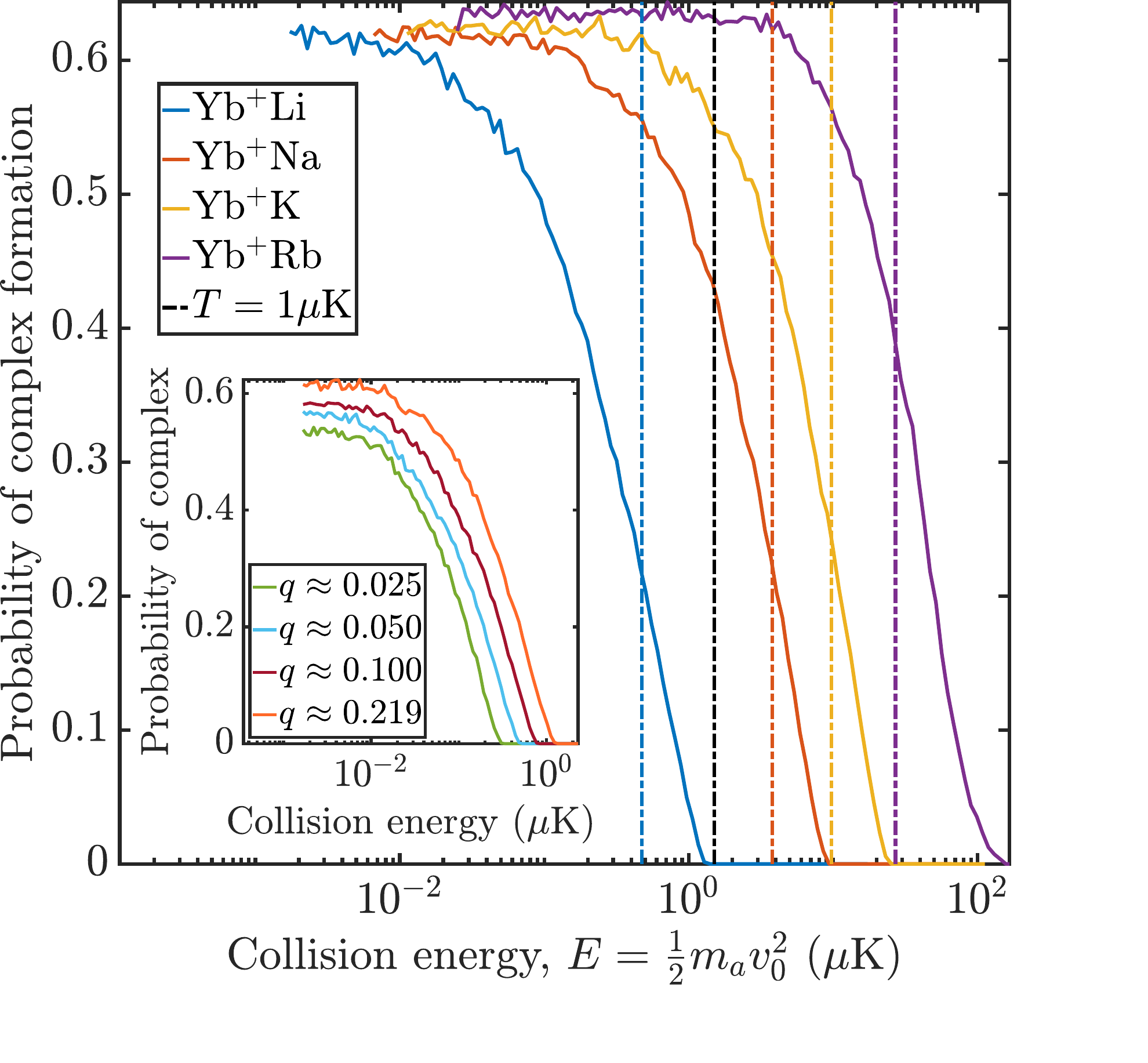}
    \caption{Probability of complex formation as a function of collision energy, for different atomic species. The four main curves show the fraction of collisions which formed a complex, for 10,000 head-on collisions of atoms with Yb$^+$, each for 100 different fixed collision energies, with $r_0=\textrm{20,000}a_0$ and $D_e=1$K. The simulation's initial rf phase $\Omega_\textrm{rf}t_0$ was uniformly randomly chosen from $[0,2\pi]$. The black vertical line is the lowest temperature used previously in the paper, $1\mu$K, corresponding to a collision energy of $1.5\mu$K via $\frac{1}{2}m_av_0^2=\frac{3}{2}k_BT$. The colored vertical lines are $W_0^\textrm{3D}$, the heating energy scale from \cite{Cetina2012}. The inset shows the probability of complex formation versus collision energy for Yb$^+$Li, for four $q_x$ values, keeping $\Omega_\textrm{rf}$ fixed.} 
    \label{fig4}
\end{figure}

Finally, we looked more closely into how the probability of complex formation depends on the collision energy for head-on collisions. Fig.~\ref{fig4} shows that as the collision energy increases, the probability of complex formation decreases, in a trend resembling a negative tanh function, consistent with the trend in fractal dimension seen in head-on atom-diatom collisions \cite{Croft2014}. The probability decreases with increasing collision energy, up to a certain threshold energy, after which complex formation is suppressed. The threshold energy is greater for heavier atoms, consistent with the fact that they form more complexes at fixed temperature (see Fig. \ref{fig3}). 

The threshold energy is consistently 2-3 times the energy scale for the minimum possible temperature of a sympathetically cooled rf-trapped ion, 
\begin{equation}
    W_0^\textrm{3D}=\frac{8}{3\pi}\left(\frac{m_a}{m_i+m_a}\right)^{5/3}\left(\frac{m_i^2\omega^4C_4}{q_x^2}\right)^{1/3},
\end{equation}
where $\omega\approx\frac{1}{2}\Omega_\textrm{rf}\sqrt{a_x+\frac{1}{2}q_x^2}$, derived by Cetina et al.~\cite{Cetina2012}. This limit arises from the micromotion-induced heating, as the ion is pulled from the center of the trap. We find that complex formation is much more likely if the initial collision energy is less than $W_0^\textrm{3D}$, so that when the atom approaches the ion before colliding, the trap can transfer an amount of energy, on the order of $W_0^\textrm{3D}$, from the atom-ion oscillation, into the ion micromotion, leading to a negative-energy temporary atom-ion bound state. In other words, if the collision energy is too high, then not enough energy can be transfered away, into the ion's heating, to form a chaotic complex. 

The inset in Fig. \ref{fig4} shows that as the Mathieu parameter increases (by increasing the rf voltage $\propto u_\textrm{rf}$, while keeping $\Omega_\textrm{rf}$ constant), the probability of complex formation increases, and the collision energy threshold increases. Also, as $q_x$ increases, $W_0^\textrm{3D}$ increases: fixing $\Omega_\textrm{rf}$ and using $a_x\ll \frac{1}{2}q_x^2$, we have $W_0^\textrm{3D}\propto q_x^{2/3}$. This gives yet another way to see the connection between heating and chaotic scattering. 

The initial rf phase, $\Omega_\textrm{rf}t_0\textrm{ mod }2\pi$, was randomized for Fig.~\ref{fig4}. If this phase were fixed, then Fig.~\ref{fig4} would show oscillations of approximately $\pm 0.1$ in the probability. This is related to the fact that complexes are more likely to form when the rf phase, at the moment of collision, is within a certain range \cite{Cetina2012}; this phase differs if the initial velocity varies, due to the different time before collision. Changing the sphere radius $r_0$ causes the stable regions shown in Fig.~\ref{fig1} to shift and stretch in ($\theta,\phi$) space, for the same reason. Thus, to obtain a reliable probability of complex formation, it is necessary to effectively average over different initial phases, as in Fig.~\ref{fig4}, or over different velocities, as in Fig.~\ref{fig3}.



In this Letter, although our system is a two-body problem, one of the bodies is continuously perturbed by a time-dependent external field, the Paul trap quadrupole field. Using GPU-accelerated MD simulations for exhaustive sampling, we conclude that the Paul trap acts as an effective third body, which exchanges energy with the two-body atom-ion subsystem and the heating of the ion. If the energy stolen by the trap, on the order of the initial heating, is greater than the atom-ion initial collision energy, they form an ion-atom complex--for a lifetime distributed according to a power law, resulting in non-hyperbolic chaotic dynamics. Furthermore, regardless of the short-range physics, colder and heavier atoms actually result in greater initial heating, by exerting a greater charge-induced-dipole force over a greater time on the ion, facilitating the action of the effective-third-body Paul trap on the ion's heating, leading to the onset of chaos in this very simple system that can be studied in the laboratory. It is worth noting that in addition to our classical treatment of the atom-trapped-ion system, another study conducted a full quantum mechanical treatment of Yb$^+$Li in one dimension, which did not show the expected spectral signatures of quantum chaos \cite{Rittenhouse2025}. The reason behind this apparent contradiction is still not clear, so more research will be dedicated to elucidating the connection between classical chaos and quantum chaos in trapped ion-atom systems. 



The authors acknowledge the support of the United States Air Force Office of Scientific Research [grant number FA9550-23-1-0202].

\end{document}